\documentclass{article}

\pdfoutput=1
\usepackage[T1]{fontenc}
\usepackage[latin1]{inputenc}
\usepackage[english,brazilian]{babel}
\usepackage{multirow}
\usepackage{graphicx}
\usepackage{color}
\usepackage{float}
\usepackage{enumitem}
\usepackage{caption}
\usepackage{url}
\usepackage[usenames,dvipsnames,svgnames,table]{xcolor}
\usepackage{tikz,pgf}
\usepackage{hyperref}
\usetikzlibrary{mindmap,shadows}

\newcommand{\Leila}{\textcolor{red}{}}
\newcommand{\Simone}{\textcolor{ForestGreen}{}}
\newcommand{\Luciana}{\textcolor{blue}{}}

\usepackage{tcolorbox}

\newcommand{\acao}[1]
          { \tcbset{colback=red!90!black,colframe=red!90!black}
             \tcbox[size=small, on line] {\textcolor{white}{\textsf{\textbf{#1}}}}
             }

\newcommand{\caixaVerde}[1]
          { \tcbset{colback=LimeGreen!90!black,colframe=LimeGreen!90!black}
             \tcbox[size=small, on line] {\textcolor{black}{{\textbf{#1}}}}
             }

\newcommand{\caixaLaranja}[1]
          { \tcbset{colback=Dandelion,colframe=Dandelion}
             \tcbox[size=small, on line] {\textcolor{black}{{\textbf{#1}}}}
             }

\newcommand{\caixaAzul}[1]
          { \tcbset{colback=Cyan!70!black,colframe=Cyan!70!black}
             \tcbox[size=small, on line] {\textcolor{black}{{\textbf{#1}}}}
             }

\newcommand{\var}[1]{\textcolor{green!75!black}{\textsf{#1}}}
\newcommand{\acaover}[1]{\textcolor{red!90!black}{#1}}

\begin{document}

\title{Entendendo o Pensamento Computacional \\ {\small (Versão Preliminar)}}
\author{Leila Ribeiro \\ \href{mailto:leila@inf.ufrgs.br}{leila@inf.ufrgs.br}
   \and Luciana Foss \\ \href{mailto:lfoss@inf.ufpel.edu.br}{lfoss@inf.ufpel.edu.br}
   \and Simone André da Costa Cavalheiro \\ \href{mailto:simone.costa@inf.ufpel.edu.br}{simone.costa@inf.ufpel.edu.br}}

\date{}

\maketitle

\begin{abstract}
O objetivo deste artigo é esclarecer o significado de Pensamento Computacional. Diferencia-se o raciocínio lógico do computacional e discute-se a importância do Pensamento Computacional na resolução de problemas.  Os três pilares do Pensamento Computacional - Abstração, Automação e Análise - são delineados, destacando-se o papel de cada um deles no desenvolvimento das habilidades necessárias para o processo de solução de problemas.\\

\end{abstract}

{
\selectlanguage{english}
\begin{abstract}
The goal of this article is to clarify the meaning of Computational Thinking. We differentiate logical from computational reasoning and discuss the importance of Computational Thinking in solving problems. The three pillars of Computational Thinking - Abstraction, Automation and Analysis - are outlined, highlighting the role of each one in developing the skills needed for the problem-solving process.
\end{abstract}
}

\section{Introdução}

Para entender o que é o \emph{pensamento computacional}, precisamos entender o que é \emph{computação}. E para entender o que é computação, a melhor maneira é um olhar  histórico, pois entendendo a origem dos conceitos podemos compreendê-los em maior plenitude. O grande objetivo da
Computação é  \emph{"raciocinar sobre o raciocínio"}.
Porém, diferente da Filosofia, aqui não estamos pensando de forma mais ampla sobre o raciocínio, mas sim interessados no processo de racionalização do raciocínio, ou seja, formalização do mesmo, o que permite a sua automação e análise (matemática).

A questão de formalização do raciocínio está intimamente relacionada à resolução de problemas. Para entender isto, tomemos como exemplo o raciocínio lógico. O objetivo do raciocínio lógico é basicamente encontrarmos (ou deduzirmos) verdades. O processo utilizado é, partido-se de premissas, que são fatos aceitos como verdades, utiliza-se regras bem definidas (do sistema lógico que se está usando) para encontrar novas verdades (conclusões). A dedução em si, que é a sequência de regras utilizadas é comumente chamada de \emph{prova} (de que a conclusão é verdadeira). O problema que está sendo resolvido é se uma sentença é ou não verdadeira: se encontrarmos uma prova a partir de sentenças que já sabemos que são verdadeiras confirmando a veracidade de uma nova sentença, ela será aceita como verdadeira. Podemos enxergar o raciocínio ou pensamento computacional como uma generalização do raciocínio lógico: um processo de transformação de entradas em saída, onde as entradas e a saída não são necessariamente sentenças verdadeiras, mas qualquer coisa (elementos de um conjunto qualquer), sendo que as entradas e a saída nem precisam ser do mesmo tipo,  e as regras que podemos utilizar não são  necessariamente as regras da lógica, mas um conjunto qualquer de regras ou instruções bem definidas. Da mesma forma que o produto do raciocínio  lógico é a prova, o produto do raciocínio computacional é a sequência de regras que define a transformação, que comumente chamamos de \emph{algoritmo}. O problema que está sendo resolvido aqui é como transformar a entrada na saída. Exemplos concretos seriam: dado um número, como encontrar seus fatores primos? Dada uma pilha de provas de alunos, como ordenar essas  provas? Dado um mapa rodoviário, como encontrar uma rota? Dados os ingredientes, como fazer um bolo? 

\begin{center}
 \includegraphics[width=.5\textwidth]{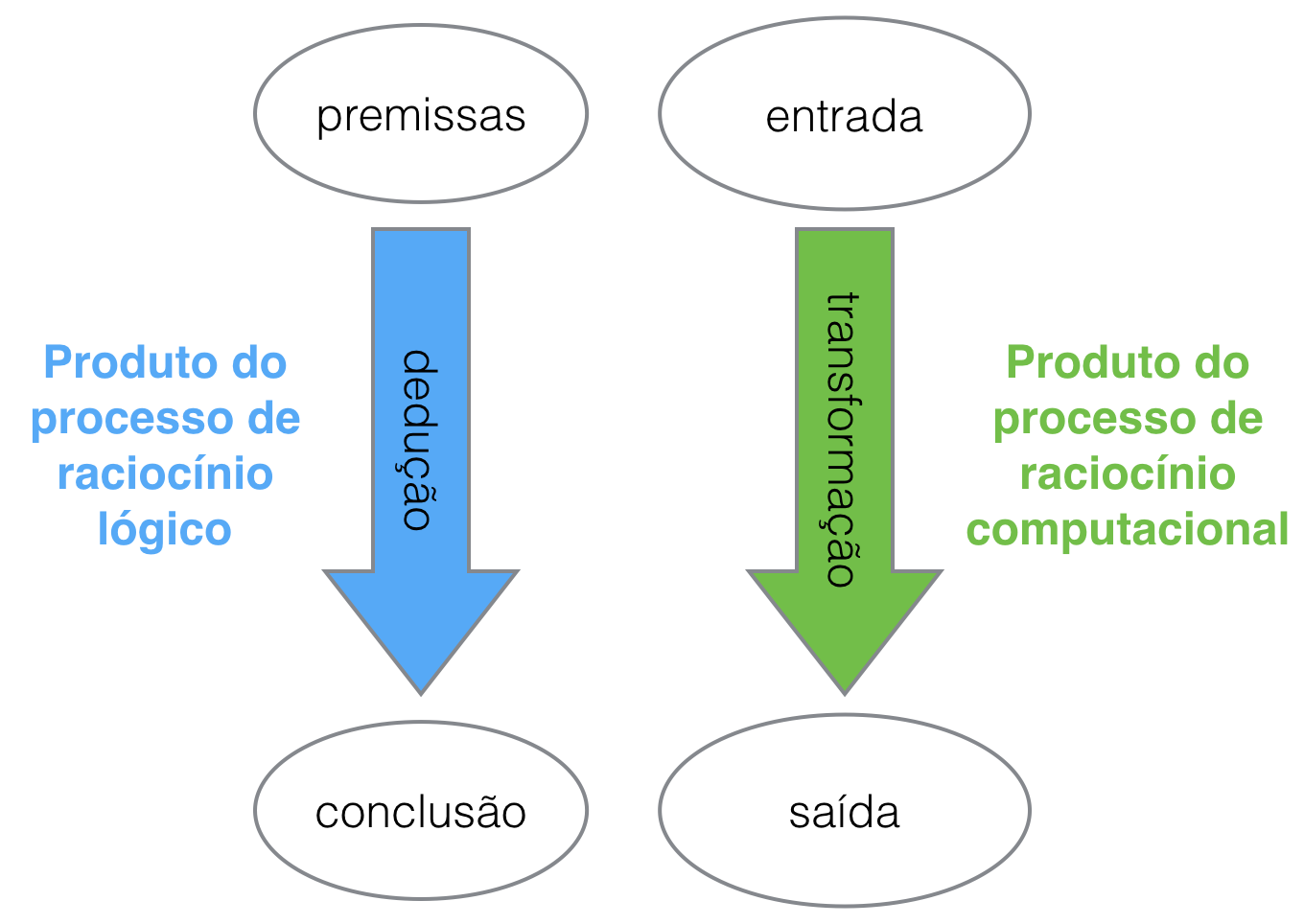}
\end{center}

Como o resultado do processo de raciocínio computacional deve ser uma descrição clara e não-ambígua de um processo,
a Computação está fortemente baseada  na Matemática, que provê uma linguagem precisa para descrição de modelos. Mas, diferente da Matemática, o objeto da Computação são os processos, ou seja, em Computação se contrói modelos de processos. Esses modelos, comumente chamados de \emph{algoritmos},  podem ser bastante abstratos, descritos em linguagem natural ou linguagens de especificação, ou programas em uma linguagem de programação.

Pode-se argumentar que na Matemática também usa-se diversas abstrações para nos ajudar a resolver problemas. Então  por que precisamos
de  Computação? Somente para automatizar a solução do problema? Não, muito mais que isso. Vamos discutir este ponto em um exemplo.

Um professor quer ensinar os alunos fatorar um número em seus fatores primos. Ele tipicamente explica os passos que os alunos devem seguir e demonstra em alguns exemplos, ou seja, o professor apresenta um algoritmo para os alunos. Por vezes, este algoritmo é, além de apresentado de forma oral, descrito em português em um livro. Os alunos, então, seguem o processo várias vezes para aprender este procedimento. Mas e se o problema fosse, ao invés da fatoração,  ordenar uma pilha de provas de alunos. Como o professor explicaria aos alunos como a tarefa deve ser realizada? E se quiséssemos que, ao invés de ordenar uma pilha dada, os alunos mesmos descrevessem como fariam a ordenação? Quais técnicas o professor utilizaria para ajudar os alunos a solucionar este problema? Note que o problema é \emph{"descreva o processo de ordenação de uma pilha de provas"}. Não é uma tarefa trivial. A Matemática não nos ajuda a resolver este tipo de problema, pois não provê as abstrações necessárias para descrever a solução. Além disso, não é objeto da Matemática
investigar   \emph{Como  construímos uma prova?}, ou mais genericamente, \emph{Como construímos um algoritmo?}. A ênfase do raciocínio  ou pensamento computacional não são apenas os produtos em si (provas ou algoritmos),  e sim o processo  de construção desses produtos, ou seja, além das abstrações necessárias para descrever algoritmos,  o pensamento computacional engloba também técnicas para a construção de algoritmos que, podem ser vistas como técnicas de solução de problemas.

\begin{figure}[!ht]
\begin{tikzpicture}
  \path[small mindmap,
    concept color=black,
    text=white,
    pensamento/.style = {concept, concept color=LimeGreen!90!black},
    tecnologia/.style = {concept, concept color=Cyan!70!black},
    cultura/.style = {concept, concept color=Dandelion}]

    node[concept, ball color=black, font=\bfseries] {Pensamento Computacional}
    [clockwise from=90]
    child[pensamento, nodes={pensamento}] {
      node  {Abstração}
      [clockwise from=150]
      child { node {Dados} }
      child { node {Processos} }
      child { node {Técnicas de construção de algoritmos}  }
        }
    child[tecnologia, nodes={tecnologia}, grow=0 ] {
      node {Análise}
      [clockwise from=60]
      child { node {Viabilidade}  }
      child { node  {Eficiência}}
      child { node {Correção} }
    }
    child[cultura, nodes={cultura}, grow=180] {
        node {Automação}
         [clockwise from=-120]
                 child  {node {Linguagensl} }
                 child  {node {Máquinas} }
                 child {node {Modelagem computacional}}
             }
;
\end{tikzpicture}
\caption{Pilares do Pensamento Computacional.} \label{fig-mindmap}
\end{figure}
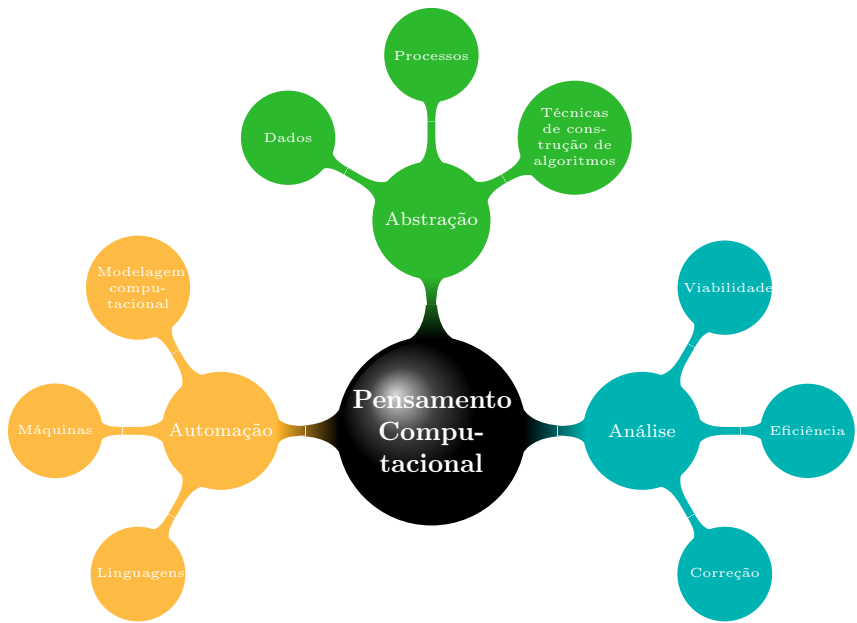

A evolução da Computação, em especial das áreas de Teoria da Computação e Engenharia de Software, descrevem a trajetória da nossa aquisição de conhecimento com relação a como sistematizar (e se possível, automatizar) o \emph{processo} de resolução de problemas.
Essa habilidade, de sistematizar, representar e analisar a atividade de resolução de problemas é chamada de \emph{raciocínio ou pensamento computacional}\footnote{
O termo \emph{pensamento computacional} é uma tradução do termo original em inglês \emph{computational thinking}. Embora do ponto de vista filosófico existam diferenças entre os termos \emph{raciocínio} e
\emph{pensamento}, neste artigo usaremos os dois termos como sinônimos.}.
A Figura~\ref{fig-mindmap} ilustra os pilares do \emph{Pensamento Computacional}, que serão detalhados nas próximas seções.

%

\normalsize

\section{Solução de Problemas e Pensamento Computacional}
\label{sec-solucao}

\newcommand{\problema}[2]
           {
           \begin{tcolorbox}[colback=green!5!white,colframe=green!75!black,title=#1]
           \textit {#2}
           \end{tcolorbox}
           }

Vamos analisar agora em mais detalhes alguns problemas e como encontrar suas soluções.
\problema{Problema 1}
 {Ordene uma pilha com 10 figurinhas em ordem crescente. Considere que não há números repetidos e os números podem variar de 1 a 10.000.}

Este problema pode ser solucionado facilmente por alunos que tenham aprendido a noção de ordem entre números. Como são apenas 10 figurinhas, normalmente eles apenas colocam todas na sua frente e vão buscando  o próximo número para montar a pilha ordenada, fazendo as comparações necessárias mentalmente.

\problema{Problema 2}
 {Ordene uma pilha com 1.000 figurinhas em ordem crescente.}

Neste caso, apesar do problema ser essencialmente o mesmo (ou seja, é apenas uma nova instância do problema anterior), a solução ad hoc descrita acima é de difícil implementação porque quando colocamos 1.000 figurinhas na mesa fica difícil visualizar  qual o próximo número. A estratégia mais usada  neste caso é, antes de ordenar, dividir a pilha de acordo com a centena, e depois cada pilha de acordo com a dezena. Ou seja, dividir o problema em problemas menores até que a solução seja quase trivial. Depois das pilhas menores estarem ordenadas, precisa-se juntá-las de forma adequada para encontrar a solução do problema. Mas esta não é a única forma de resolver este problema, existem muitas outras, algumas mais e outras menos eficientes.

\problema{Problema 3}{Descreva como ordenar uma pilha com figurinhas em ordem crescente.}

Agora o problema é como descrever o método de ordenação, ou seja, o algoritmo utilizado para ordenar,  para que o processo possa ser replicado, seguido por outras pessoas. Esta tarefa é bem mais difícil do que as anteriores, principalmente porque para descrever o processo de ordenação, precisamos falar sobre estruturas de dados (neste caso, uma pilha) e usar operações para definir o que deve ser feito em cada passo. Mas sem ter formalizados os conceitos de estruturas de dados e operações para definir processos (um algoritmo é uma descrição de um processo) é difícil solucionar este problema. Saber dar instruções de forma clara e precisa é uma habilidade muito necessária para todas as pessoas, e esta habilidade requer treinamento adequado. Além dos conhecimentos sobre os dados e como construir processos, precisa-se ter domínio da linguagem na qual os processos serão descritos. E a linguagem a ser usada depende de quem executará o processo: se for uma pessoas, pode-se usar linguagem natural, se for um computador, deve-se usar uma linguagem de programação. A grande diferença normalmente está no nível de abstração, pois linguagens naturais tendem a ser mais abstratas. Em breve discutiremos em mais detalhes a questão das linguagens.

\problema {Problema 4} {O processo de ordenação descrito pode ser executado mais rápido se houverem mais pessoas para ajudar? Qual o número de pessoas seria o ideal?
Existe alguma forma mais eficiente de ordenar as figurinhas? Dadas duas estratégias de ordenação, qual a melhor?
O algoritmo descrito está correto, ou seja, no final da execução, a pilha está ordenada?}

Solucionar este problema envolve analisar o algoritmo, ou método de ordenação, descrito na solução do Problema 3. Mas, mesmo que este método esteja descrito de forma precisa, a análise da correção e eficiência do método não é uma tarefa trivial, apesar de ser de extrema importância porque um algoritmo que não gera o resultado desejado é inútil, bem como um que gera o resultado esperado, mas que demoraria demais para gerar este resultado (dependendo do problema e instância considerada, uma  solução pode demorar vários anos para ser encontrada e esperar pode não ser viável do ponto de vista prático).

\begin{figure}[htb]
\begin{center}
 \includegraphics[width=1\textwidth]{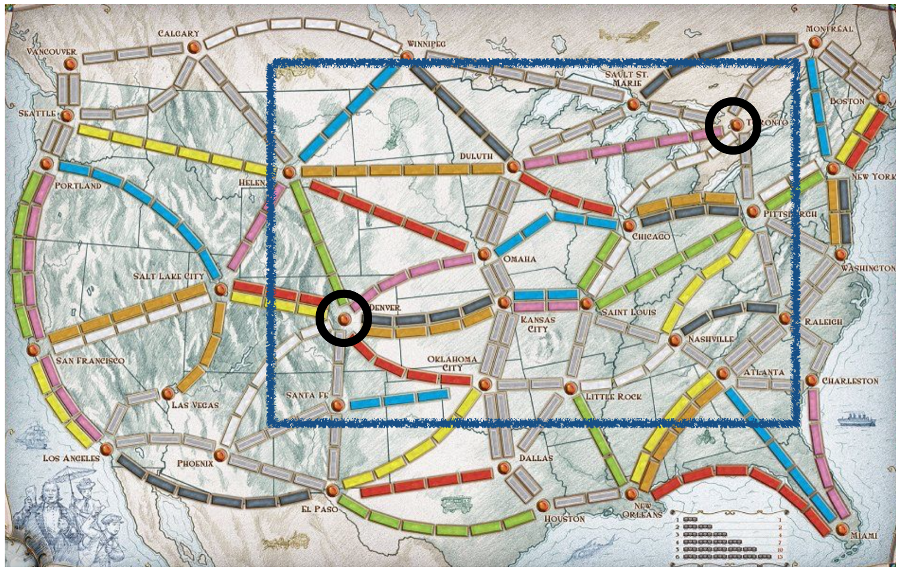}
\end{center}
\caption{\label{fig-mapa} Mapa}
\end{figure}

\problema{Problema 5}   {Dado o mapa de cidades da Figura  \ref{fig-mapa}, encontrar o menor caminho entre Denver e Toronto.}
Este parece um problema simples, mas para dar uma ideia do tamanho do problema, somente considerando as cidades de dentro do quadrado azul, e sem repetir nenhuma cidade no caminho, existem 1.360 caminhos diferentes entre Denver e Toronto!

Como estes problemas, existem inúmeros outros que são do nosso cotidiano, e que às vezes precisamos resolver, sem termos tido na nossa
 formação um ferramental que nos auxilie nesta tarefa (por exemplo os problemas 6, 7 e 8 a seguir).

\problema{Problema 6} {Descreva como encontrar o menor caminho entre duas cidades de um mapa.}

\problema{Problema 7}  {Dados um conjunto de professores, com as disciplinas que cada um pode ministrar;
um conjunto de turmas de cada disciplina a serem ministradas;
o conjunto de salas disponíveis; e
restrições de horários de professores, turmas e salas,
elabore uma alocação de professores a turmas e turmas a salas.}

\problema{Problema 8}   {Dada uma mala com um volume máximo e um conjunto de ítens a serem colocados na mala, cada um com um valor representando a importância dele ser levado na mala e um volume, como escolher quais devem ser colocados na mala de forma a  maximizar o valor da mala, sem exceder seu volume?}

\begin{figure}[htbp]
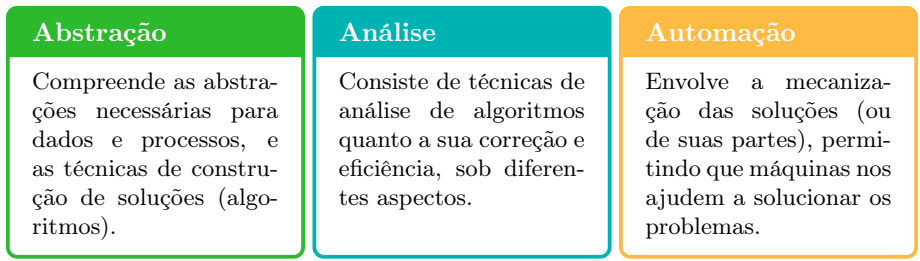

\tcbset{width=(\linewidth-2mm)/3,nobeforeafter,arc=1mm,colback=white,fonttitle=\bfseries,fontupper=\small, left=2mm,right=2mm,top=1mm,bottom=1mm,equal height group=parbox}

\noindent
\begin{tcolorbox}[colframe=LimeGreen!90!black,parbox,adjusted title={Abstração}]
  Compreende  as abstrações necessárias para dados e processos, e as técnicas de construção de soluções (algoritmos).
\end{tcolorbox}\hfill%
\begin{tcolorbox}[colframe=Cyan!70!black,parbox=false,adjusted title={Análise}]
 Consiste de técnicas de análise de algoritmos quanto a sua correção e eficiência, sob diferentes aspectos.
\end{tcolorbox}\hfill%
\begin{tcolorbox}[colframe=Dandelion,parbox=false,adjusted title={Automação}]
Envolve a mecanização das soluções (ou de suas partes), permitindo que máquinas nos ajudem a solucionar os problemas.
\end{tcolorbox}\hfill%

\tcbset{reset}
\caption{\label{fig-pilares} Pilares do Pensamento Computacional}
\end{figure}

\bigskip
Os pilares do Pensamento Computacional \cite{Wing3717}, mostrados na Figura \ref{fig-pilares}, provêm as habilidades necessárias para resolver os problemas citados anteriormente.

\begin{figure}[htbp]
    \begin{tcolorbox}[colback=white,colframe=white!75!black,title=(a) Linguagem Visual]
     \includegraphics[width=1\textwidth]{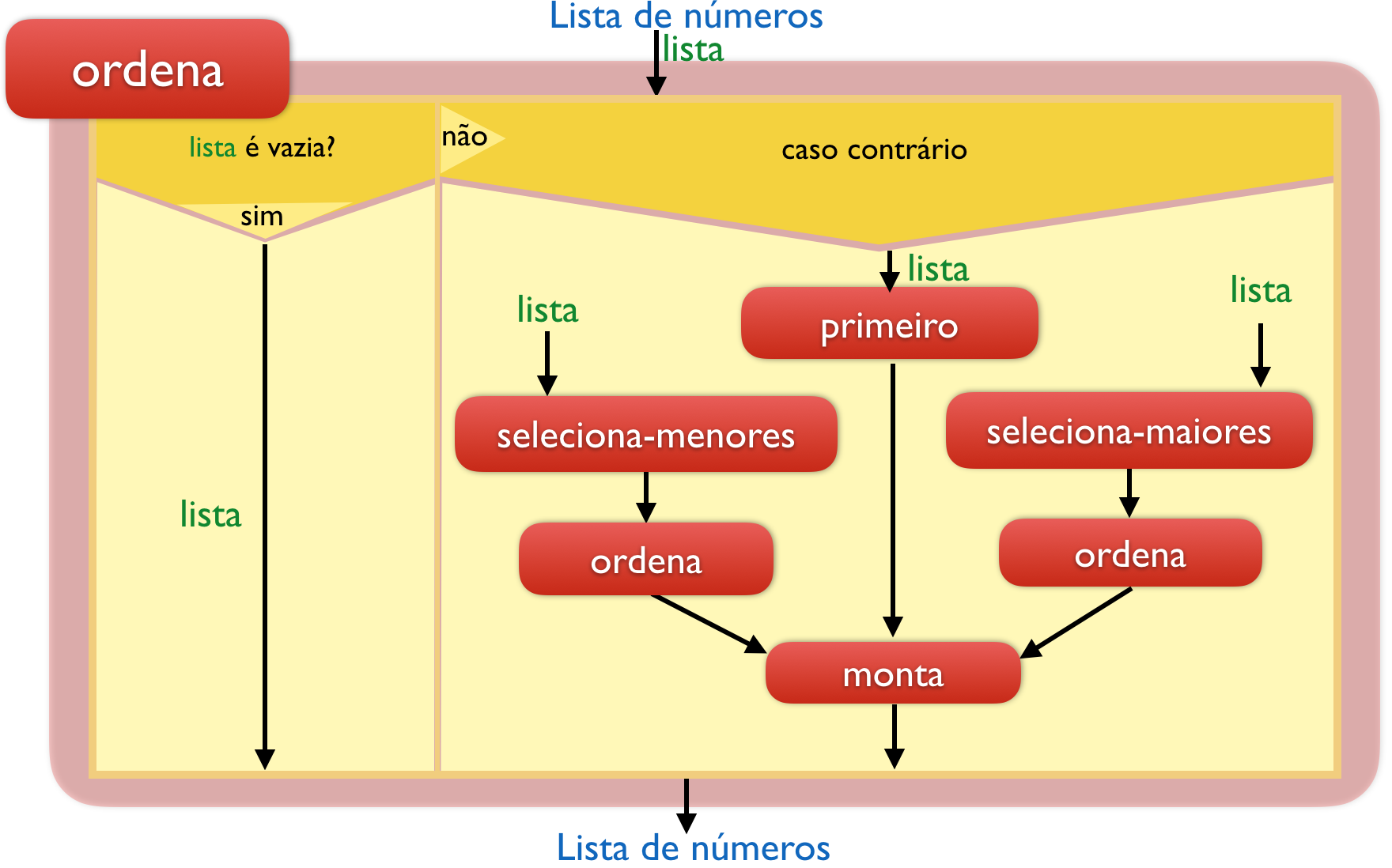}
    \end{tcolorbox}

 \begin{tcolorbox}[colback=white,colframe=white!75!black,title=(b) Linguagem Natural]
\begin{tcolorbox}[colback=black!5!white,colframe=black!75!white]
\begin{minipage}{15cm}
\sf
\begin{tabbing}
Se \= a \var{lista} for vazia \\
      \> então  devolver a própria \var{lista}\\
      \> senão \= \acaover{montar} uma lista contendo: \\
                                              \>  \> a lista dos números \acaover{menores} que o primeiro, \acaover{ordenada};  \\
                                              \>  \> o \acaover{primeiro} elemento; e \\
                                               \>  \> a lista dos números \acaover{maiores} que o primeiro, \acaover{ordenada}.

\end{tabbing}
\end{minipage}
\end{tcolorbox}
 \end{tcolorbox}

 \begin{tcolorbox}[colback=white,colframe=white!75!black,title=(c) Linguagem de Programação]
\begin{tcolorbox}[colback=black!5!white,colframe=black!75!white]
\begin{minipage}{15cm}
\sf
\begin{tabbing}
(define (\acaover{ordena} \var{lista})\\
  (cond \=\\
          \>  [(\acaover{vazia?} \var{lista})   \var{lista}]\\
          \>  [else  (\acaover{monta} \=\\
                                              \>  \>  (\acaover{ordena} (\acaover{seleciona-menores} \var{lista}));  \\
                                              \>  \>  (\acaover{primeiro} \var{lista}); e \\
                                               \>  \> (\acaover{ordena} (\acaover{seleciona-maiores} \var{lista})))]))

\end{tabbing}
\end{minipage}
\end{tcolorbox}
\end{tcolorbox}

\caption{\label{fig-ordena} Descrição do algoritmo \textsf{Ordena}}
\end{figure}

 A seguir, vamos dar 3  exemplos de soluções para o problema de ordenar uma lista de números em ordem crescente (uma abstração do problema de ordenar figurinhas) usando 3 linguagens diferentes: uma visual, linguagem natural e linguagem de programação. O algoritmo apresentado é um algoritmo tradicional de ordenação chamado \emph{quicksort} \cite{quicksort}. A Figura \ref{fig-ordena}(a)  mostra o algoritmo através de um diagrama. As setas representam o fluxo dos dados, as caixas vermelhas representam ações e a parte amarela pontos de decisão. A ideia é que a \var{lista} (a lista de números) entra na  caixa \acao{ordena}, e então é testado se a \var{lista} está vazia ou não. Em caso positivo, o  algoritmo termina retornando a própria \var{lista} vazia (que está ordenada pois não contém nenhum elemento). Em caso negativo, é identificado o primeiro elemento da \var{lista} (ação \acao{primeiro}), e a \var{lista} é dividida em 2 partes, uma contendo os elementos menores que o primeiro (ação \acao{seleciona-menores}) e outra contendo os elementos
 maiores que o primeiro  (ação \acao{seleciona-maiores}). Cada uma destas sublistas resultantes são ordenadas (repetindo o processo \acao{ordena} para cada uma) e no final o resultado é construído juntando-se essas sublistas ordenadas (ação \acao{monta}). A linguagem visual é interessante porque deixa evidente o fluxo e as ações envolvidas.  Este mesmo processo pode ser descrito em língua portuguesa (Figura \ref{fig-ordena} (b)). Note que a descrição do algoritmo em português é apenas uma frase, que foi quebrada em linhas diferentes na figura para maior clareza. É uma frase simples, mas descreve de forma sucinta e precisa o processo que deve ser seguido para ordenar a lista. Quando temos uma descrição precisa da solução, a implementação em uma linguagem de programação pode ser imediata: a Figura \ref{fig-ordena} (c) mostra como ficaria o programa correspondente descrito em uma linguagem funcional (\textsf{Racket} \cite{racket}). Basicamente, o programa tem os mesmos elementos principais que a descrição textual, mas com uma sintaxe mais enxuta e rígida.
 Isso exemplifica um dos pontos que queremos enfatizar neste texto: \emph{programar é fácil, o difícil é saber construir a solução dos problemas}. Se soubermos construir uma frase (ou texto) preciso em português que descreve um processo,  a programação é um simples trabalho de tradução. Claro, dependendo da linguagem de programação utilizada a tradução pode ser mais fácil ou difícil, pois linguagens de programação diferentes oferecem abstrações diferentes, mas ainda assim é uma tradução, a questão que realmente exige maior
 esforço e conhecimento é a construção da solução em si. E é este o foco do \emph{Pensamento Computacional}.

Em 2006, Wing \cite{Wing:2006:CT:1118178.1118215} utiliza o termo \emph{Pensamento Computacional } para apresentar a visão de que todas as pessoas podem se beneficiar do ato de pensar como um cientista da Computação. Informalmente, o pensamento computacional \cite{Wing:2011:theLink} descreve a atividade mental envolvida na formulação de problemas para admitir soluções computacionais e na proposta de  soluções.  As soluções (algoritmos) podem ser executadas por seres humanos ou máquinas, ou de maneira mais geral, por combinações de seres humanos e máquinas.

Já, em 1962, Alan Perlis \cite{Perlis:1962:computer} argumentava que todos deveriam aprender a programar computadores no nível universitário. Ele identificou que a execução automatizada dos processos, explorada pela programação, mudaria a forma como os profissionais de todas as áreas pensariam sobre seu trabalho. No contexto da educação básica, na década de 1980, Papert \cite{Papert:1980:MCC:1095592} introduziu e popularizou a ideia de que computadores e o pensamento procedural poderiam afetar o modo como as crianças pensam e aprendem. Ao desenvolver o construcionismo (uma abordagem do construtivismo), defendia que o uso do computador (ou de ferramentas similares) na educação permitiria ao estudante desenvolver o seu raciocínio na  solução de problemas  e construir o seu próprio conhecimento.


O desenvolvimento do \emph{Pensamento Computacional} não tem como objetivo direcionar as pessoas a pensarem como computadores. Ao contrário, sugere que  utilize-mos a nossa inteligência, os fundamentos e os recursos da computação para abordar os problemas. Importante também observar que raciocinar computacionalmente é mais do que programar um computador. A Sociedade Internacional de Tecnologia em Educação (ISTE) e a Associação de Professores de Ciência da Computação (CSTA) \cite{CTtoolkit} operacionalizaram o termo \emph{Pensamento Computacional} como um processo de resolução de problemas que inclui: formular problemas de uma maneira que seja possível usar um computador e outras ferramentas para ajudar a resolvê-los; organizar e analisar dados de maneira lógica; representar dados através de abstrações; descrever soluções através do pensamento algorítmico (uma série de passos ordenados);  identificar, analisar e implementar possíveis soluções com o objetivo de alcançar a combinação mais eficiente e eficaz de etapas e recursos; generalizar e transferir este processo de resolução de problemas para uma grande variedade de problemas.

Segundo Wing \cite{Wing:2006:CT:1118178.1118215}, o \emph{Pensamento Computacional} pode ser colocado como uma das habilidades intelectuais básicas de um ser humano, comparada à ler, escrever, falar e fazer operações aritméticas. Habilidades estas que servem para descrever e explicar situações complexas. Nesta linha de raciocínio, o \emph{Pensamento Computacional} é mais uma linguagem (junto com as linguagens escrita e falada, e a matemática) que podemos usar para falar sobre o universo e seus processos complexos.

\section{Abstração}

A abstração é um mecanismo importante no processo de solução de problemas, o qual permite simplificar a realidade e representar os aspectos mais relevantes de um problema e sua solução. Ela compreende os seguintes aspectos:

\begin{description}
\item [\caixaVerde{Dados}]: Abstrações que permitem descrever as informações envolvidas na solução de um problema (dados de entrada e saída);
\item [\caixaVerde{Processos}]: Abstrações que permitem definir os algoritmos que descrevem a solução de um problema, as quais devem estar de acordo com a capacidade de compreensão do leitor;
\item [\caixaVerde{Técnicas de Construção de Algoritmos}]: Técnicas que permitem obter a solução de problemas complexos de forma mais simples.

\end{description}

\subsection{Abstrações para representar Informações}

Em Matemática, um dos conceitos mais fundamentais é \emph{Número}, 	que é uma abstração para quantidades.
Várias áreas da Matemática usam esta noção, como a Álgebra, a Geometria e a Probabilidade.
Já a Lógica se baseia em outro tipo de noção: a noção de \emph{Conjunto}. 

Para descrever algoritmos, que tipo de noções são necessárias? Claro, vai depender do que os algoritmos fazem, mas se um algoritmo representa uma transformação de recursos (dados de entrada) em resultados (dados de saída), precisamos ser capazes de representar esses recursos e resultados de alguma forma. Como os algoritmos devem ser genéricos (ou seja, funcionar para várias entradas diferentes), a entrada e a saída devem ser representadas por conjuntos de elementos. Dependendo da finalidade do algoritmo, os elementos podem ser muito simples (um número, por exemplo), ou complexos (uma pilha de provas de alunos, um mapa, uma ficha de paciente de hospital, etc). Para podermos descrever algoritmos necessitamos poder falar sobre esses dados, sejam eles simples ou complexos. E para isso precisamos das abstrações adequadas. Quando a entrada é um número ou uma palavra, podemos usar os conhecimentos já adquiridos durante anos em Matemática ou Português. Mas quando queremos processar uma pilha de provas para, por exemplo, ordenar de alguma forma, precisamos usar uma abstração para esta pilha. Quando queremos descrever como se encontra uma rota em um mapa rodoviário, precisamos falar do mapa e como ele é organizado para poder explicar para alguém como se procura um caminho. A diferença entre um número e uma pilha de provas é que o número representa um conceito indivisível,  uma quantidade, enquanto a pilha é composta por unidades menores, que são as provas. E cada prova, por sua vez, pode ser composta de uma coleção de informações (nome do aluno, questões, respostas, nota). E para explicar como ordenar esta pilha de provas, precisamos acessar cada elemento da pilha, bem como as informações contidas em cada prova.
Para descrever como encontrar uma rota em um mapa precisamos enxergar o mapa não como uma unidade, mas como um conjunto de cidades ligadas por estradas. Ou seja, para descrever algoritmos nós precisamos enxergar dados como composições de dados mais simples. Assim, temos vários níveis de abstração que podem ser usados para resolver um problema: no caso das provas, podemos enxergar a pilha como um todo, ou selecionar uma das provas, ou pegar uma informação de uma das provas. O nível de abstração escolhido dependerá do que se quer realizar em cada passo do algoritmo. Mas precisamos entender e poder falar sobre todos.

As abstrações  de dados mais importantes em Computação são  \Leila:

\begin{itemize}
\item Registros: um registro representa  uma coleção de informações de um objeto. Por exemplo, um registro de prova pode conter nome do aluno, questões, respostas, nota, etc, registros podem ser usados também para descrever dados de carteiras de identidade, formulários, cartão de respostas do vestibular, etc;
\item Listas: uma lista é uma sequência de dados. Listas podem ser usadas como abstração para pilha de provas, baralho de cartas, cadeias de DNA, lista de compras, fila de banco, partituras (listas de notas musicais), etc;
\item Grafos: um grafo é uma estrutura que contém entidades (chamadas vértices) e relacionamentos (chamados arcos). Grafos podem ser usados para representar uma infinidade de estruturas, como redes sociais, mapas, árvores genealógicas, etc.
\end{itemize}

Essas abstrações precisam ser trabalhadas de forma concreta e depois formalizadas, da mesma forma que o conceito de número na Matemática, para permitir que os alunos tenham capacidade de trabalhar sobre elas depois.

\subsection{Abstrações para descrever Algoritmos \Leila }

Além de abstrações para dados, precisamos de técnicas para descrever as soluções em forma de algoritmos. Um algoritmo é composto por instruções que devem ser executadas de uma forma e na ordem definida para se atingir a solução desejada. Portanto, para se definir um algoritmo é necessário saber quais as instruções básicas que se pode usar, e quais operações podem ser usadas para se montar descrições dos procedimentos a partir dessas instruções básicas.

As instruções básicas dependem de quem vai ler o algoritmo. Se o leitor já sabe como ordenar uma lista, a instrução \emph{``Ordene a lista''} é adequada. Caso contrário, precisa-se definir melhor como realizar esta instrução, através de instruções mais básicas que o leitor consiga entender. Em linguagens de programação, as instruções básicas são os comandos pré-definidos da linguagem, e existem bibliotecas de instruções que podem ser utilizadas. Para construir as soluções dos problemas para os quais não existem instruções básicas  que os
resolvam, usam-se operações que combinam instruções básicas de maneira a definir processos mais elaborados.
Essas operações são basicamente de 3 tipos:

\begin{itemize}
\item Composição: permite juntar vários passos na descrição de um algoritmo. Esses passos podem ser conectados de várias formas diferentes (sequencial, paralela, por dependências, etc);
\item Escolha: permite definir pontos de escolha em um algoritmo, que são momentos de decisão nos quais o próximo passo a ser executado depende da situação atual do processo;
\item Repetição: permite que ações sejam repetidas em um algoritmo, de forma controlada. Existem várias formas de se definir como as repetições devem ser executadas (por exemplo, laços ou  recursão).
\end{itemize}

Essas operações são implementadas de diversas formas em diferentes linguagens de programação. Um algoritmo é, portanto, uma combinação de instruções usando operadores de composição, escolha e repetição.

\subsection{Técnicas para Construir Algoritmos}
Para se construir um algoritmo, não basta conhecer as abstrações de dados e processos. São necessárias técnicas que nos permitem chegar
com mais facilidade do enunciado de um problema a uma solução. Entre estas técnicas, destacam-se:
\begin{description}
\item [Decomposição]: É a técnica mais importante para se solucionar um problema, e consiste em decompor o problema em problemas menores, solucioná-los e combinar as soluções para obter a solução do problema original;
\item [Generalização]: É uma técnica que consiste em construir uma solução (algoritmo) mais genérico a partir de outro, permitindo que este novo algoritmo seja utilizado em outros contextos. Reutilizar e adaptar algoritmos é fundamental, e exige um grande poder de abstração. Muitas vezes problemas que, a primeira vista parecem totalmente diferentes podem ser solucionados pelo mesmo algoritmo fazendo-se apenas pequenas modificações. Programas ou algoritmos são descrições de procedimentos, portanto, podem ser usados como dados para outros programas ou algoritmos. Essa noção de que programas são dados, chamada de \emph{ meta-programação},   é fundamental e permite que se construam soluções extremamente elegantes, genéricas e simples para problemas complexos.;
\item [Transformação]: A técnica de transformação consiste em utilizar a solução de um problema para solucionar outro, através de transformação. Essas transformações podem ser feitas em diferentes contextos:
para  utilizar um algoritmo já existente para resolver o problema (reuso);
para realizar melhorias em uma solução existente (refinamento);
para adaptar soluções existentes a outras realidades (evolução);
para compreender as relações entre problemas (redução); etc.
\end{description}

\section{Automação \Luciana}

A abstração nos permite encontrar e descrever um modelo de solução para um problema, e a automação é a mecanização de todas ou parte das tarefas da solução para resolver o problema usando computadores.

Para podermos automatizar a solução de um problema, primeiro é necessário saber se essa automatização é possível.
Nem todos os problemas podem ser resolvidos com o uso de computadores, existem vários problemas que não são passíveis de mecanização, chamados não-computáveis.
Em alguns casos, apenas parte da solução pode ser executada por um computador. Por exemplo, não existe algoritmo que pode determinar se duas funções são equivalentes, mas é possível, dada uma entrada, verificar se duas funções produzem a mesma saída. Outros problemas não-computáveis são: determinar se um conjunto de dominós pode cobrir um tabuleiro; determinar se um algoritmo sempre termina; determinar se uma equação (polinomial) sempre tem uma solução (inteira); verificar se um programa tem vulnerabilidades de segurança, etc.

A automação envolve diferentes aspectos que devem ser levados em conta:
\begin{description}
\item [\caixaLaranja{Máquina}]: Escolha da máquina (computador) a ser utilizada para automatizar a solução de problema;
\item [\caixaLaranja{Linguagem}]: Escolha da linguagem de programação a ser utilizada para descrever a solução;
\item [\caixaLaranja{Modelagem Computacional}]: Utilização de modelos que simulam o comportamento de sistemas reais e permitem validar a solução de um problema.
\end{description}

Para que a mecanização seja possível, o computador deve ser capaz de interpretar as abstrações do modelo. Nesse contexto, um computador poderia ser um dispositivo mecânico, elétrico ou biológico (como por exemplo o DNA ou computadores moleculares) com capacidade de processamento, armazenamento e comunicação. Ou também poderia ser um humano, que segue fielmente os passos de um algoritmo, realizando o processamento de informações de forma mecânica.
É importante saber escolher qual o tipo de computador (ou combinação de computadores) é o mais adequado para realizar uma tarefa desejada. Por exemplo, para preencher uma nota fiscal de venda, é melhor o vendedor preencher manualmente e fazer os cálculos no papel? Ou preencher manualmente e fazer os cálculos usando uma calculadora? Ou ainda, preencher usando um aplicativo de computador que fará os cálculos de forma automática? Para fazer a escolha adequada, é importante que se conheçam as características de cada máquina: para que elas servem, qual a dificuldade de utilizá-las, que tipos de problemas elas podem apresentar, como resolver esses problemas, etc.

Escolhido o computador adequado, deve-se traduzir a solução do problema (algoritmo) para uma linguagem compreendida pelo computador.
Cada tipo de computador reconhece uma (ou várias) linguagem(ns)  diferente(s). Por exemplo, um computador tradicional compreende dados e instruções representadas por sequências de zeros e uns; o DNA compreende informações compotas por sequências de bases A (adenina), C (citosina), T (tinina) e G (guanina); já um humano compreende sentenças descritas em diferentes linguagens naturais (português, inglês, espanhol, etc) e também linguagens formais como a matemática. Apesar de serem mais facilmente compreendidas, as linguagens naturais nem sempre são a melhor opção quando se quer descrever de forma precisa nossas abstrações. Uma linguagem natural é essencialmente ambígua e subjetiva, o que permite diferentes interpretações para uma mesma instrução.
Já as linguagens compreendidas pelos computadores tradicionais, como PCs, notebooks, tabletes, celulares, entre outros, não sofrem desse tipo de problema. Essas linguagens (linguagens de máquina) usam uma representação binária bastante precisa. Deste modo, qualquer informação ou instrução deve ser codificada por sequências de zeros e uns, que são reconhecidas pelo computador e determinam as ações que ele deve realizar. Além disso, dependendo do tipo de arquitetura do computador, as instruções compreendidas por ele também podem variar. Assim, para um indivíduo conseguir descrever a solução de seu problema para que um computador a compreenda e a execute, ele deveria conhecer a linguagem de máquina do computador escolhido.

A tarefa de descrever procedimentos usando linguagem de máquina é bastante árdua, visto que essa codificação é bastante distante da linguagem natural e dependente da máquina escolhida. O ideal, portanto, seria podermos descrever nossos procedimentos em uma linguagem mais próxima à natural e que o computador pudesse compreendê-la. Esse é exatamente o papel das linguagens de programação de mais alto nível. Essas linguagens geralmente são mais próximas às linguagens naturais e independentes da arquitetura dos computadores. As instruções destas linguagens possuem um nível de abstração maior, isto é, elas geralmente correspondem a uma sequência de instruções de uma linguagem de máquina. Contudo, uma linguagem de programação não pode ser interpretada diretamente pelos computadores e, portanto, existem traduções dessas para as linguagens de máquina. Assim, dizemos que as linguagens de programação possuem um nível de abstração maior do que as linguagens de máquina. Mesmo entre as linguagens de programação, existem diferentes níveis de abstração. Por exemplo, a linguagem Java possui um nível de abstração maior do que a linguagem C, assim como as linguagens funcionais têm maior nível de abstração do que as linguagens procedurais. A escolha da linguagem a ser utilizada, deve levar em conta essa característica. Quanto maior o nível de abstração, maior é a facilidade de descrever o algoritmo (solução do problema) nesta linguagem. 

A utilização de modelos pode auxiliar no entendimento de um problema, permitindo a simulação do comportamento dos sistemas envolvidos, bem como de soluções propostas. Os modelos podem ser físicos ou matemáticos. Por exemplo, pode-se construir modelos físicos de pontes que permitem medir deformações sofridas por tais estruturas ao receberem uma determinada carga; ou ainda, pode-se construir modelos matemáticos que podem ser simulados com o uso de um computador. A modelagem computacional fornece recursos para tratar problemas complexos e que envolvem um elevado número de variáveis. Para isso, é proposto o uso de métodos numéricos para tratamento do problema, associados a ferramentas computacionais e técnicas avançadas de programação. Exemplos  de  simulação  computacional podem ser encontrados nas mais diversas áreas como, no estudo de sistemas biológicos, no desenvolvimento de projetos e jogos, na previsão metereológica, etc.

O mercado está repleto  de  ambientes  para  simulação  computacional  que  disponibilizam diversos recursos para construção de modelos de sistemas reais e a decisão de qual e como utilizar precisa ser tomada. Para isso, algumas considerações devem ser feitas: Como validar um modelo? Como tratar os resultados obtidos de uma simulação? Como saber se os resultados são válidos para o sistema real?

\section{Análise \Simone}

A  Ciência da Computação provê fundamentos teóricos sólidos  e uma rica teoria
para análise e classificação de problemas, permitindo descobrir se um problema tem ou não solução computacional,
e também se pode ter algoritmo eficiente que o resolva, antes mesmo de tentar construir o algoritmo.
A análise é de extrema importância pois fundamenta argumentação crítica sobre os problemas e suas soluções (algoritmos).
De forma geral, a análise pode ser de 3 tipos:

\begin{description}
\item [\caixaAzul{Viabilidade}]: Análise da viabilidade de se encontrar uma solução computacional para o problema;
\item [\caixaAzul{Correção}]: Verificação se o algoritmo construído é mesmo a solução desejada para o problema em questão;
\item [\caixaAzul{Eficiência}]: Avaliação da eficiência do algoritmo, sob vários aspectos.
\end{description}

A viabilidade já foi discutida na seção anterior, quando discutiu-se que nem todos problemas tem solução computacional.
Um exemplo clássico de problema não-computável é o problema de escrever um programa que determina se outros programas param ou entram num laço (processam indefinidamente). Quer-se um programa que possa ler o código de qualquer outro programa e seus respectivos dados e que possa retornar se o processo pára ou se executa indefinidamente um conjunto de instruções. Pode até parecer que este programa é viável  (e até mesmo fácil) de ser construído, mas na verdade ele não pode existir. A razão pela qual ele é inviável é que este programa poderia ser fornecido como entrada dele mesmo (já que ele é um programa e a entrada dele é qualquer programa). A partir deste ponto, é bastante fácil construir um paradoxo afirmando que, se o  programa pára, então ele deve entrar num laço e se ele entra num laço, então ele deve parar.

Retornando aos problemas que possuem soluções computacionais, como saber se um algoritmo proposto resolve o problema em questão? Uma solução está ``correta'' quando funciona exatamente como se espera em todas as situações. Se a solução foi dada por um programa de computador, afirma-se que o programa está ``correto'' quando ele fornece a saída esperada para todo valor possível de entrada. A questão é que o conjunto de todas as entradas possíveis para um programa, exceto para casos triviais, é extremamente grande. Ademais, os problemas que hoje em dia se apresentam nas mais diversas áreas do conhecimento possuem, em geral, soluções complexas. Concomitantemente, muitas ferramentas e metodologias surgiram para auxiliar neste processo.

Simulações e testes são algumas das técnicas utilizadas para encontrar erros e avaliar se os programas possuem características desejadas. Elas envolvem a execução de partes do programa ou de todo o sistema para avaliar propriedades de interesse, como por exemplo, se o programa responde corretamente a determinadas entradas, se executa as principais funções dentro de um tempo aceitável, se atinge o resultado geral esperado, entre outros. Um conjunto de testes pode ser bom para avaliar uma determinada propriedade (por exemplo, funcionalidade), mas ineficaz para avaliar outras características (por exemplo, eficiência). Algumas métricas são definidas para um bom conjunto de testes, tais como: ele deve incluir as condições iniciais e as sequências de entrada para as simulações; deve especificar as saídas esperadas; deve incluir uma descrição de sua finalidade ou do requisito que está em análise (ou ambos); entre outros. No entanto, em geral, nenhum conjunto de testes será bom e eficaz para todo o tipo de análise.

Outra técnica de análise consiste na definição e utilização de modelos matemáticos para simular e verificar sistemas reais. A partir de uma especificação precisa (modelo matemático) é possível construir uma prova formal (utilizando de argumentação lógica e técnicas de demonstração) que garante que o modelo satisfaz determinadas propriedades. Diversas metodologias também já foram propostas para demonstrar que o projeto de um sistema está correto com respeito à sua especificação. Refinamentos (transformações precisas) podem ser especificados para transformar um modelo matemático num projeto e um projeto numa implementação, a qual, ao final do processo, é correta por construção.

Além da correção, outro ponto a ser analisado em um algoritmo é sua eficiência, permitindo avaliar e comparar diferentes algoritmos quanto ao uso de recursos como tempo, memória, processador, energia, comunicação, etc.

Vamos supor que se precise avisar um colega que a reunião da tarde foi cancelada. Poderia-se  enviar uma mensagem pelo celular para o colega. Alternativamente, poderia-se telefonar para ele. Também se poderia enviar um e-mail ou mesmo, ir pessoalmente a sua casa. Enfim, há diversas alternativas para avisá-lo do cancelamento da reunião.
Qual deve ser utilizada? Para escolher, pode-se levar em consideração os recursos disponíveis (por exemplo, celular, telefone), o tempo que será necessário para avisar usando cada alternativa, o custo de cada alternativa, etc.
Da mesma forma, existem diversos algoritmos que resolvem um mesmo problema. Para escolher qual utilizar em cada situação, precisamos poder analisar quantitativamente
os algoritmos para dar subsídio ao processo de escolha da melhor alternativa.

Retomando o problema da ordenação.  Supõe-se que queremos colocar uma lista de números em ordem crescente, como na seção \ref{sec-solucao}. Existe uma grande variedade de soluções possíveis (algoritmos) para este problema. A ordenação por inserção é um processo que funciona da maneira como muitas pessoas ordenam as cartas em um jogo de baralho. Ao receber as cartas viradas na mesa, pegam-se as cartas, uma a uma, inserindo-as na posição correta. Para encontrar a posição correta, o que se faz é comparar a carta pega da mesa com cada uma das cartas que já estão na mão, da direita para a esquerda. Outro método de ordenação bastante rápido e eficiente é a ordenação
que foi descrita na seção \ref{sec-solucao}, Figura \ref{fig-ordena}.
Se considerarmos que recebemos os elementos em uma ordem aleatória e que temos as mesmas ferramentas para resolver o problema, a ordenação por inserção é considerada uma solução eficiente para um pequeno número de elementos. Já se considerarmos quantidades maiores de elementos, a ordenação da Figura \ref{fig-ordena} passa a ser mais rápida. Ou seja, existem soluções mais eficientes do que outras para resolver um mesmo problema. Essas diferenças em eficiência ficam mais evidentes quando são consideradas instâncias grandes dos problemas (neste caso, listas com muitos elementos).

Existem problemas  para os quais não se encontrou até o momento soluções computacionais eficientes. Para estes problemas,  chamados intratáveis, existem soluções (isto é, existem algoritmos que o resolvem), mas na prática levariam tanto tempo para chegar a um resultado (por vezes anos ou séculos, dependendo do tamanho da instância do problema) que se tornam inúteis. Já foi mostrado que muitos problemas de grande interesse prático se enquadram nesta categoria. Um exemplo é o problema de encontrar  o caminho mais curto entre duas cidades em um mapa. Este problema pode ser resolvido  calculando todas as rotas possíveis e comparando-as para identificar a melhor rota. Porém, isso se torna completamente inviável pois, em geral, mesmo em instâncias com poucas  cidades, há um número enorme de rotas a considerar e o algoritmo poderia levar milhões de anos para ser dar uma resposta. É por este motivo que os programas que existem hoje para determinar rotas em mapas não necessariamente determinam a melhor rota para um motorista, e sim uma lista de rotas que podem ser encontradas de forma rápida. É importante saber como identificar se um problema é intratável,  para que não se tente encontrar solução eficiente para um problema que já foi classificado como intratável. Muitas vezes, pequenas modificações no enunciado do problema podem torná-lo tratável computacionalmente.

Importante destacar que para a maioria absoluta dos programas (problemas) reais não existe uma única técnica de análise que permita afirmar que o sistema (ou a solução) está livre de qualquer erro, que satisfaz todos os requisitos e propriedades desejados e que vai operar sempre conforme o esperado.  Para garantir a qualidade do sistema, uma combinação de técnicas e ferramentas deve ser utilizada, o que requer treinamento e capacitação dos desenvolvedores. De maneira similar, mesmo que não se queira construir programas, uma boa análise das soluções dadas para muitos problemas do dia-a-dia depende da sistematização dos procedimentos e de treinamento da argumentação lógica. E uma fundamentação consistente para atingir este objetivo pode ser encontrada na Ciência da Computação. 

\bibliographystyle{plain}
\bibliography{bibliografia}

\end{document}